\documentclass[10pt,conference]{IEEEtran}
\IEEEoverridecommandlockouts
\usepackage{algpseudocode}
\usepackage{algorithm}
\usepackage{color}
\usepackage{bm}
\usepackage{amsfonts}
\usepackage{amssymb}
\usepackage{amscd}
\usepackage{graphics}
\usepackage[dvips]{graphicx}
\usepackage{epsfig}
\usepackage{subfigure}
\usepackage{caption}
\usepackage[cmex10]{amsmath}
\usepackage{array}
\usepackage{eqparbox}
\usepackage{fancyhdr}
\usepackage{enumerate}
\usepackage{stackrel}
\usepackage{pdflscape}
\usepackage{longtable}
\usepackage{lineno}
\usepackage{mathtools}
\usepackage{breqn}
\newcommand{\ie}{{\em i.e.}}
\newcommand{\etc}{{\em etc.}}

\newcommand{\iid}{i.i.d.}
\newcommand{\apriori}{{\em a priori}}

\newcommand{\secref}[1]{Section~\ref{#1}}
\newcommand{\figref}[1]{Fig.~\ref{#1}}

\newcommand{\noteref}[1]{Note~\ref{#1}}

 \DeclarePairedDelimiter{\ceil}{\lceil}{\rceil}

\newcommand{\abs}[1]{\left\vert#1\right\vert}

\makeatletter
\def\blfootnote{\xdef\@thefnmark{}\@footnotetext}
\makeatother

\newtheorem{theorem}{Theorem}[section]

\newtheorem{note}[theorem]{Note}

\newcommand{\qed}{\nobreak \ifvmode \relax \else
      \ifdim\lastskip<1.5em \hskip-\lastskip
      \hskip1.5em plus0em minus0.5em \fi \nobreak
      \vrule height0.75em width0.5em depth0.25em\fi}

\def\BibTeX{{\rm B\kern-.05em{\sc i\kern-.025em b}\kern-.08em
    t\kern-.1667em\lower.7ex\hbox{E}\kern-.125emX}}

\usepackage{newlfont}
\begin{document}
\title{Energy-efficient Decision Fusion for Distributed Detection in Wireless Sensor Networks}
\author{\IEEEauthorblockN{N. Sriranga, K. G. Nagananda, R. S. Blum, \emph{Fellow IEEE}, A. Saucan and P. K. Varshney, \emph{Life Fellow IEEE}}\thanks{N. Sriranga, K. G. Nagananda and R.S. Blum are with the Department of Electrical  Engineering, Lehigh University, Bethlehem, PA 18015, USA. E-mail: \texttt{nandansriranga@gmail.com}, \texttt{kgnagananda@gmail.com}, \texttt{rblum@ece.lehigh.edu}. A. Saucan and P. K. Varshney are with the Department of Electrical  Engineering and Computer Science, Syracuse University, Syracuse, NY 13244, USA. E-mail: \texttt{\{asaucan, varshney\}@syr.edu}. This material is based upon work partially supported by the U. S. Army Research Laboratory, the U. S. Army Research Office under grant number W911NF-17-1-0331 and grant number W911NF-14-1-0339.}}
\maketitle
\thispagestyle{empty}
\pagestyle{empty}  

\begin{abstract}
This paper proposes an energy-efficient counting rule for distributed detection by ordering sensor transmissions in wireless sensor networks. In the counting rule-based detection in an $N-$sensor network, the local sensors transmit binary decisions to the fusion center, where the number of all $N$ local-sensor detections are counted and compared to a threshold. In the ordering scheme, sensors transmit their unquantized statistics to the fusion center in a sequential manner; highly informative sensors enjoy higher priority for transmission. When sufficient evidence is collected at the fusion center for decision making, the transmissions from the sensors are stopped. The ordering scheme achieves the same error probability as the optimum unconstrained energy approach (which requires observations from all the $N$ sensors) with far fewer sensor transmissions. The scheme proposed in this paper improves the energy efficiency of the counting rule detector by ordering the sensor transmissions: each sensor transmits at a time inversely proportional to a function of its observation. The resulting scheme combines the advantages offered by the counting rule (efficient utilization of the network's communication bandwidth, since the local decisions are transmitted in binary form to the fusion center) and ordering sensor transmissions (bandwidth efficiency, since the fusion center need not wait for all the $N$ sensors to transmit their local decisions), thereby leading to significant energy savings. As a concrete example, the problem of target detection in large-scale wireless sensor networks is considered. Under certain conditions the ordering-based counting rule scheme achieves the same detection performance as that of the original counting rule detector with fewer than $N/2$ sensor transmissions; in some cases, the savings in transmission approaches $(N-1)$. 
\end{abstract} 
\vspace{0.1in}

\section{Introduction} \label{sec:intorduction}
Energy-efficient wireless sensor networks (WSNs) have drawn much attention owing to limited computational capabilities of low-cost and low-power sensors that are distributed randomly in an environment to collect observations and make decisions \cite{Sadler2005}. For example, optimal sensor selection (bit allocation) schemes for distributed detection with constrained system resources such as minimum communication have appeared in \cite{Gini1998} - \nocite{Yu1998}\nocite{Yu1998a}\nocite{Kasetkasem2001}\cite{Hu2001}. Energy-efficient routing protocols, cluster-based ad hoc routing scheme for multi-hop sensor networks, and passive clustering techniques have been proposed to enhance the lifetimes of the energy-constrained sensors; see, \cite{Raghunathan2002} - \nocite{Intanagonwiwat2003}\nocite{Gharavi2003}\cite{Kwon2003}. Other techniques for minimizing energy consumption in WSNs include distributed data compression and transmission \cite{Baek2004}, and collaborative signal processing \cite{Kumar2002}.

In the context of statistical inference in WSNs, different sensor transmission policies have been devised to achieve energy efficiency. This is especially relevant in multi-sensor systems, wherein the local sensors transmit their decisions to a fusion center where the received information is combined to yield the global inference \cite{Varshney1997}. For instance, in the censoring sensors scheme, the sensors transmit only highly informative observations ({\ie}, when their likelihoods are very large or very small) to the fusion center; otherwise, the transmissions are ceased or ``censored'' \cite{Rago1996} - \nocite{Patwari2003}\nocite{Appadwedula2005}\nocite{Marano2006}\cite{Appadwedula2008}. Another approach involves ordering sensor transmissions \cite{Blum2008}, where a sensor transmits at a time inversely proportional to the magnitude of its likelihood ratio; when it transmits, it transmits the likelihood ratio. When sufficient evidence is collected at the fusion center for decision making, the transmissions are stopped. The ordering scheme incurs the same error probability as the optimum energy-unconstrained approach (where all the sensors transmit) \cite{Poor1994}, but with far fewer sensor transmissions thereby yielding significant energy (time) savings. Based on \cite{Blum2008}, numerous papers on ordered transmissions for inference in WSNs have appeared in the literature \cite{Blum2008a} - \nocite{Rawas2011}\nocite{Blum2011}\nocite{Braca2011}\nocite{Braca2012}\nocite{Xu2012}\nocite{Hesham2012}\nocite{Marano2013}\nocite{Ayeh2013}\cite{Sriranga2018}. A shortcoming of the ordering scheme is that, the sensor likelihood ratios are transmitted in unquantized form which utilizes more energy and communications bandwidth of the network.

In \cite{Niu2004} (see also \cite{Niu2005a} - \nocite{Niu2005}\nocite{Niu2006}\cite{Niu2008}), the counting rule detector was proposed for distributed detection in large-scale WSNs with efficient utilization of the network's communication bandwidth. Rather than transmitting the unquantized versions, the local-sensor decisions are transmitted to the fusion center in binary form (1 denotes detection, 0 otherwise). This minimizes the overall communication bandwidth utilization in the network. At the fusion center, the number of local-sensor detections are counted and compared to a threshold to make a system-level inference. A drawback of this approach is that the fusion center has to collect the binary decisions of \emph{all} the sensors in order to make the system-level decision. This results in heavy utilization of the communications bandwidth which is undesirable especially in large-scale WSNs. Secondly, in WSNs with randomly deployed sensors, since the sensors that are located far away from the target are likely to incur higher probability of error because their received signal strength will be very weak due to attenuation. The local-decision errors will degrade the overall system-level detection performance. 

In this paper, the underlying principle of ordering sensor transmissions proposed in \cite{Blum2008} is exploited to reduce the number of sensor transmissions in the counting rule detector of \cite{Niu2004}, thereby improving its communications or bandwidth efficiency. As an example, the problem of target detection in a distributed setting in large-scale WSNs considered in \cite{Niu2006} is revisited. The setup comprises a stationary target, a large number $N$ of randomly deployed local sensors and a fusion center in the region of interest (ROI). The signal power emitted by the target is assumed to follow the isotropic power attenuation model, {\ie}, the signal power of a sensor's observation is inversely proportional to the distance between that sensor and the target (which can be exploited to order sensor transmissions). Since the received signal strength at a sensor is inversely proportional to the distance between the sensor and the target (due to power attenuation), it is highly likely that the sensors closer to the target are more informative about the target's presence/absence than those that are far away. 

Therefore, a sensor is programmed to transmit at a time inversely proportional to a function of its observed signal. Note that, the distance between a sensor and the target is unknown, so the transmission policy depends only on the sensor's observation. It is assumed that each sensor uses an orthogonal channel to transmit to the fusion center. When sufficient evidence is collected at the fusion center, the transmissions are stopped. It is also assumed that the fusion center has sufficient computational capability, and that there are feedback channels from the fusion center to all the sensors in the WSN. This feedback mechanism is important for improving the energy efficiency of the WSN and has been employed by previous works on this topic; see, for example, \cite{Blum2008}, \cite{Blum2011}, \cite{Sriranga2018}. For large $N$ and strong signal power, the average number of transmissions saved (ANTS) approaches $N/2$, while achieving the same system-level detection probability as the counting rule detector of \cite{Niu2004}. Simulation results show that significant ANTS is achieved even when the target's emitted power is not very large. Under certain conditions (which will be made precise later) the ANTS approaches $(N-1)$. 

The remainder of the paper is organized as follows. The system model for target detection in large-scale WSNs is presented in \secref{sec:model}. A brief review of the counting rule detector and ordering sensor transmissions is presented in \secref{sec:review}. The ordering-based counting rule detector is developed in \secref{sec:ocrd}. Numerical results and discussions are presented in \secref{sec:simulations}. \secref{sec:conclusion} concludes the paper. 

\section{System model} \label{sec:model}
\begin{figure}
\centering
\includegraphics[height=3.25in,width=3.75in]{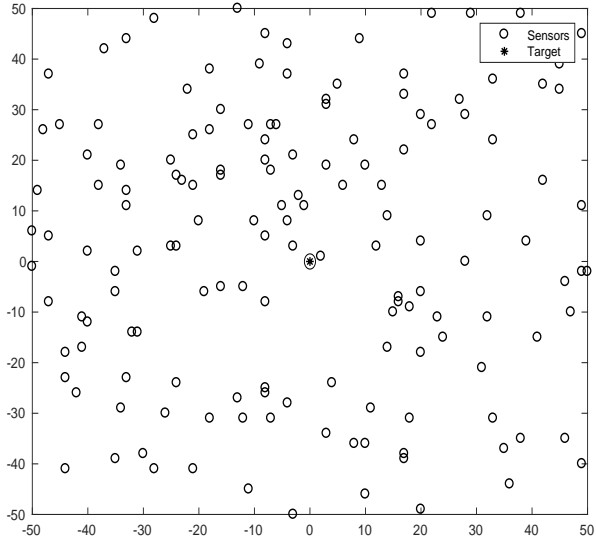}
\caption{Distribution of sensors in the ROI. The target is located at the point $(0, 0)$.}
\label{fig:roi}
\end{figure}

The WSN model considered in this paper is adopted from the one in \cite{Niu2006}. The target whose presence/absence is to be detected is located at $(x_t, y_t)$ in the two-dimensional plane, which constitutes the ROI represented by a square of area $b^2$ as shown in \figref{fig:roi}. A total of $N$ sensors are randomly deployed in this ROI, such that the locations of sensors are unknown to the WSN. The sensor locations are assumed to be {\iid} and follow a uniform distribution in the ROI. If $(x_i, y_i)$ denote the coordinates of $i^{\text{th}}$ sensor ($i = 1, \dots, N$), then the PDF of its location is given by 
\begin{align}
p(x_i, y_i) = 
\begin{cases}
\frac{1}{b^2},~ \frac{-b}{2} \leq x_i, y_i \leq \frac{b}{2}, \\
0,~\text{otherwise}.
\end{cases}
\label{eq:pdf_location}
\end{align}
The $i^{\text{th}}$ sensor observes $z_i$ and seeks to resolve the following binary hypothesis testing problem:
\begin{eqnarray}
\begin{aligned}
H_0: z_i &=& w_i, \\
H_1: z_i &=& s_i + w_i,
\label{eq:hypotheses}
\end{aligned}
\end{eqnarray}
where $w_i$ is the sensor receiver noise and is assumed to follow the standard Gaussian distribution, {\ie}, $w_i \sim \mathcal{N}(0, 1)$, and $s_i$ denotes the signal amplitude recorded at the sensor. The signal attenuation model is the same as the one in \cite[Section 2]{Niu2006}, where it is assumed that the signal power at a sensor decays with increasing distance from the target. More precisely, 
\begin{eqnarray}
s_i^2 = \frac{P_0}{1 + \alpha d^n_i},
\label{eq:signal_energy}
\end{eqnarray}
where $P_0$ is the power emitted by the target at distance zero, $n$ is the signal decay exponent (takes values between 2 and 3), $\alpha$ is an adjustable constant (larger $\alpha$ implies faster signal power decay), and $d_i$ is the distance between the target and the $i^{\text{th}}$ sensor given by 
\begin{eqnarray}
d_i = \sqrt{(x_i - x_t)^2 + (y_i - y_t)^2},
\end{eqnarray}
Note that, in practical applications the distance $d_i$ is unknown, since the sensors are randomly deployed. This signal attenuation model can be easily extended to three-dimensional problems. 

Each of the $N$ local sensors use the same threshold $\tau$ to make a local decision. This has been shown to be  to asymptotically optimal for the decentralized binary hypotheses testing problem \cite{Tsitsiklis1988}. The probabilities of false alarm and detection of the $i^{\text{th}}$ sensor are given, respectively, by 
\begin{align}
p_{\text{fa}_i} = p_{\text{fa}} = \int_{\tau}^{\infty} \frac{1}{\sqrt{2\pi}} e^{-\frac{t^2}{2}} dt = Q(\tau), \label{eq:local_pfa} \\
p_{\text{d}_i} = \int_{\tau}^{\infty} \frac{1}{\sqrt{2\pi}} e^{-\frac{(t - s_i)^2}{2}} dt = Q\left(\tau - \sqrt{\frac{P_0}{1 + \alpha d^n_i}}\right), \label{eq:local_pd}
\end{align}
where $Q(\cdot)$ is the complementary distribution function of the standard Gaussian given by 
\begin{eqnarray}
Q(x) = \int_{x}^{\infty} \frac{1}{\sqrt{2\pi}} e^{-\frac{t^2}{2}} dt.
\end{eqnarray}
The system level (fusion center) probabilities of false alarm and detection are denoted by $P_{\text{fa}}$ and $P_{\text{d}}$, respectively.             

\section{Review of energy-efficient sensor transmission schemes}\label{sec:review}
In this section, a brief review of the counting rule detector of \cite{Niu2004} and ordering sensor transmissions approach of \cite{Blum2008} for energy-efficient signal detection in WSNs in presented. 

\subsection{Chair-Varshney fusion rule \cite{Chair1986}}\label{subsec:chair_varshney}
The optimal Chair-Varshney decision fusion rule for distributed detection has the following test statistic \cite{Chair1986}:
\begin{eqnarray}
\lambda_1 = \sum_{i=1}^{N} \left[I_i \ln \left(\frac{p_{\text{d}_i}}{p_{\text{fa}_i}}\right) + (1 - I_i)\ln \left(\frac{1 - p_{\text{d}_i}}{1 - p_{\text{fa}_i}}\right) \right].
\label{eq:chair_varshney}
\end{eqnarray}
In \eqref{eq:chair_varshney}, $p_{\text{fa}_i}$ can be calculated using \eqref{eq:local_pfa} as long as the threshold $\tau$ is known. However, in the context of random sensor fields considered this paper, it is difficult to calculate $p_{\text{d}_i}$ from \eqref{eq:local_pd}, since it depends on the distance $d_i$ between the $i^{\text{th}}$ sensor and the target which is assumed unknown. 

\subsection{Counting rule detector \cite{Niu2004}}\label{subsec:count}
To alleviate the shortcoming of the Chair-Varshney decision fusion, the counting rule was proposed in \cite{Niu2004} (see also \cite{Niu2005a} - \nocite{Niu2005}\nocite{Niu2006}\cite{Niu2008}), where the sensors transmit their respective decisions to the fusion center in binary form. The decision of the $i^{\text{th}}$ sensor is denoted by $I_{i} \in \{1, 0\}$. $I_i = 1$ when there is detection; otherwise, it takes the value $0$. It is assumed that each sensor uses an orthogonal channel with negligible error rates to transmit its decision to the fusion center. (The assumption of orthogonal channels is commonly used in many wireless communication systems like TDMA, FDMA, {\etc}) The counting rule detector structure is given by 
\begin{eqnarray}
\lambda_2 = \sum_{i=1}^{N} I_i \stackrel[H_0]{H_1}{\gtrless} T,
\label{eq:count}
\end{eqnarray}
where $T$ denotes the test threshold. The system level performance measures ($P_{\text{fa}}$ and $P_{\text{d}}$) are given by the following expressions. For large $N$, $P_{\text{fa}}$ is given by \cite[Section 4.1]{Niu2006}
\begin{eqnarray}
P_{\text{fa}} \approxeq Q\left(\frac{T - Np_{\text{fa}}}{\sqrt{Np_{\text{fa}}(1 - p_{\text{fa}})}}\right).
\label{eq:count_Pfa}
\end{eqnarray}
The following notation is introduced before providing an expression for the system-level $P_{\text{d}}$ \cite[Section 4.2]{Niu2006}: 
\begin{eqnarray}
\gamma \!\!\!\! &=& \!\!\!\!  Q\left(\tau - \sqrt{\frac{P_0}{1 + \alpha \left(\frac{\sqrt{2}b}{2}\right)^n}}\right), \label{eq:gamma_Pd} \\
\bar{p}_{\text{d}} \!\!\!\! &=&\!\!\!\! \frac{2\pi}{b^2} \int\limits_{0}^{b/2} Q\left(\tau - \sqrt{\frac{P_0}{1 + \alpha r^n}}\right) r dr +  \left(1 - \frac{\pi}{4} \right) \gamma, \label{eq:count_pdbar} \\
\nonumber \bar{\sigma}^2 \!\!\!\! &=&\!\!\!\! \frac{2\pi}{b^2} \int\limits_{0}^{b/2} \left(1 - Q\left(\tau - \sqrt{\frac{P_0}{1 + \alpha r^n}}\right) \right)\times \\&& Q\left(\tau - \sqrt{\frac{P_0}{1 + \alpha r^n}}\right) r dr +  \left(1 - \frac{\pi}{4} \right) \gamma(1 - \gamma), \label{eq:count_sigmabar} 
\end{eqnarray}
Finally, the system-level $P_{\text{d}}$ is given by 
\begin{eqnarray}
P_{\text{d}} = Q\left(\frac{T - N\bar{p}_{\text{d}}}{\sqrt{N \bar{\sigma}^2}}\right).
\label{eq:count_Pd}
\end{eqnarray}
The optimality of the counting rule detector was proved in \cite[Section 4.5]{Niu2006}. In the next subsection, an approach is presented that will significantly improve  the efficiency of the counting rule detector, without compromising on its system-level detection probability. 

\subsection{Ordering sensor transmissions \cite{Blum2008}}\label{sec:ee_order}
In the ordering scheme to achieve energy efficiency presented in \cite{Blum2008}, the $j^{\text{th}}$ sensor transmits at a time proportional to $1/\abs{\ln (L_j)}$, where $\ln(L_j) = \ln[f_{Z_j}(z_j|H_1)/f_{Z_j}(z_j|H_0)]$ is the log-likelihood ratio. If the $j^{\text{th}}$ sensor transmits, it transmits $\ln(L_j)$ to the fusion center. The fusion center decides in favor of $H_1$ if 
\begin{align}
\sum_{j=1}^{k}\ln(L_{[j]}) > \ln\left(\frac{1-p}{p}\right) + (N-k)\abs{\ln (L_{[k]})},
\label{eq:order_H1}
\end{align}
or, it decides in favor of $H_0$ if 
\begin{align}
\sum_{j=1}^{k}\ln(L_{[j]}) < \ln\left(\frac{1-p}{p}\right) - (N-k)\abs{\ln(L_{[k]})},
\label{eq:order_H0}
\end{align}
where $L_{[j]}$ is the $j^{\text{th}}$ largest likelihood ratio, $k$ is the number of sensors that have transmitted till a given time and $L_{[k]}$ is the last sensor likelihood ratio transmission prior to that same time, $p = \text{Pr}(H_1)$, $\ln((1-p/p))$ is the test threshold for the optimal energy unconstrained Bayes detector \cite{Poor1994} (where all $N$ sensors transmit). When one of the thresholds is exceeded, the fusion center informs the sensors to stop transmission via the feedback channel. As shown in \cite[Theorem 2]{Blum2008}, if there is a distance measure $s$ (for example, mean-shift) between the competing distributions such that 
\begin{align}
\text{Pr}(\ln(L_j) > 0|H_1) \xrightarrow{s \rightarrow \infty} 1,  \\
\text{Pr}(\ln(L_j) < 0|H_0) \xrightarrow{s \rightarrow \infty} 1,
\label{eq:order_H0}
\end{align}
then ANTS $\geq N/2$ for equal priors ($p = 0.5$), without loss in error probability compared to the optimal energy unconstrained detector. A drawback of the ordering scheme is that, the local sensors transmit their log-likelihood ratio $\ln(L_j)$ which needs more energy and communication bandwidth than what is required to transmit binary decisions. The censoring sensors approach of \cite{Rago1996} is similar to ordering, and the local decisions are transmitted to the fusion center as raw data [$\ln(L_j)$]. However, censoring incurs a higher error probability compared to the energy unconstrained optimal detector. 

The method proposed in this paper achieves the same ANTS ($\geq N/2$) compared to the ordering scheme of \cite{Blum2008}; under certain conditions, the ANTS even approaches $(N-1)$. Further, since the local sensor decisions are transmitted in binary form, the energy and bandwidth savings are higher than that of the ordering scheme, while at the same time achieving the same detection probability as that of the counting rule detector of \cite{Niu2004}. In the following, it will be shown that, a slight modification of the ordering scheme of \cite{Blum2008} significantly improves the energy efficiency of the counting rule detector of \cite{Niu2004} without compromising on its detection performance. 

\section{Ordering-based counting rule}\label{sec:ocrd}
From \eqref{eq:signal_energy}, it is evident that the received signal strength at the $i^{\text{th}}$ local sensor is inversely proportional to the distance $d_i$ between the sensor and the target. That is, it is highly likely that the observations from the sensors that lie closer to the target are more informative about the target's presence or absence than those that are far away. Therefore, the $i^{\text{th}}$ sensor transmits at a time inversely proportional to a function of its received signal. Essentially, the $i^{\text{th}}$ sensor transmits at time $\frac{1}{\abs{z_i - \tau}}$, where $\tau$ is calculated using \eqref{eq:local_pfa}. By doing so, it is high likely that the sensors closer to the target will enjoy higher priority for transmission compared to those that are farther away. 

Similar to the counting rule detector of \cite{Niu2004}, the decision statistic of the local sensor is denoted by $I_i \in \{+1, 0\}$. $I_i = 1$ when there is detection; otherwise, it takes the value $0$. Note that, this scheme does not require system-level {\apriori} knowledge of the distance $d_i$ between the sensor and the target (which, in any case, is unknown). It is assumed that each sensor uses an orthogonal channel with negligible error rates to communicate with the fusion center. In addition, there is a feedback channel with a negligible error rate from the fusion center to every sensor. When sufficient evidence is collected at the fusion center for decision making ({\ie}, when a stopping criterion is satisfied), the fusion center informs the sensors via the feedback channel to stop transmission. In principle, this approach is similar to the ordering scheme of \cite{Blum2008}, however, unlike \cite{Blum2008} the local sensors transmit their decisions in binary form leading to improved energy savings. In the following, the mechanism to stop sensor transmissions is devised. 

At a given time, let $k$ denote the number of sensors that have transmitted their respective binary decision to the fusion center; therefore, $(N-k)$ denotes the total number of sensors that have not transmitted until that time. The fusion center informs (via the noiseless feedback channels) the sensors to stop transmission if one of the following two conditions is satisfied:
\begin{eqnarray}
\sum_{i=1}^{k}I_{[i]} \!\!\!\! &>& \!\!\!\! T, \label{eq:upper} \\
\sum_{i=1}^{k}I_{[i]} \!\!\!\! &<& \!\!\!\! T - (N-k), \label{eq:lower} 
\end{eqnarray}
where $I_{[i]}$ denotes the binary decision of the $i^{\text{th}}$ sensor corresponding to the highest signal power. In other words, the set $\{I_{[1]}, \dots, I_{[k]} \}$ denotes the ordered sequence of binary digits, where $I_{[1]}$ is the decision of the local sensor with the highest signal power, $I_{[2]}$ corresponds to the sensor with the second highest signal power and so on. For a fixed system-level $P_{\text{fa}}$ and local sensor-level $p_{\text{fa}}$, the threshold $T$ can be calculated using \eqref{eq:count_Pfa}:
\begin{align}
T = Q^{-1}(P_{\text{fa}})\sqrt{Np_{\text{fa}}(1 - p_{\text{fa})}} + N p_{\text{fa}}.
\label{eq:fusion center_T}
\end{align}

\begin{note}
Let $r$ denote the likelihood of the target's presence in the ROI.  The likelihood $r$ is essential to characterize the ANTS. However, this likelihood should not be confused with the prior probability that is used in the Bayesian detection criterion. The local sensors and the fusion center employ the Neyman-Pearson setting to make decisions, so knowledge of the likelihood $r$ is not involved in the detection process; it is used only for characterizing the number of transmissions saved using the proposed scheme. 
\label{note:prior}
\end{note}

It can be shown that under certain conditions, {\ie}, for large $N$ and when the likelihood of the target present is 0.5, the ANTS using the proposed scheme approaches $N/2$. Consider the case where the upper threshold is exceeded, {\ie}, the condition \eqref{eq:upper} is satisfied. Define
\begin{align}
k^{\ast} \!\! = \!\! \min_{1\leq k \leq N} \left\{\sum_{i=1}^{k}I_{[i]}  >  T  \right\}.
\label{eq:ants1_upper}
\end{align}

For convenience of analysis, and without loss of generality, $\ceil{T}$ is rounded off to the next highest integer. The ANTS for the ordering-based counting rule detector is given by 
\begin{eqnarray}
\nonumber N^{\ast} \!\!\!\! &=& \mathbb{E}[N - k^{\ast}] = \sum_{k=1}^{N} (N-k) \text{Pr}(k^{\ast} = k) \\
\nonumber &\stackrel{(a)}=&\!\!\!\! \sum_{k=1}^{\ceil{T}-1} (N-k) \text{Pr}(k^{\ast} = k) +  (N-\ceil{T}) \text{Pr}(k^{\ast} = \ceil{T}) \\ \nonumber && + \sum_{k=\ceil{T}+1}^{N} (N-k) \text{Pr}(k^{\ast} = k) \\ 
\nonumber &\stackrel{(b)}\geq&\!\!\!\! \sum_{k=1}^{\ceil{T}-1} (N-k) \text{Pr}(k^{\ast} = k) +  (N-\ceil{T}) \text{Pr}(k^{\ast} = \ceil{T}) \\
\nonumber &\stackrel{(c)}\geq& (N-\ceil{T}) \text{Pr}(k^{\ast} \leq \ceil{T})  \\
\nonumber &\stackrel{(d)}=& (N-\ceil{T}) \text{Pr}\left\{\sum_{i=1}^{\ceil{T}}I_{[i]}  \geq  T  \right\} \\
\nonumber &\stackrel{(e)}=& (N-\ceil{T})\left[\text{Pr}\left\{\sum_{i=1}^{\ceil{T}}I_{[i]}  \geq  T|H_1  \right\} r \right. \\ \nonumber && \left. +  \text{Pr}\left\{\sum_{i=1}^{\ceil{T}}I_{[i]}  >  T|H_0  \right\}(1-r) \right] \\
\nonumber &\stackrel{(f)}\geq& (N-\ceil{T})\text{Pr}\left\{\sum_{i=1}^{\ceil{T}}I_{[i]}  \geq  T|H_1  \right\} r. 
 \label{eq:ants2}
\end{eqnarray}
In going from $(a)$ to $(b)$, a positive term is dropped. In going from $(b)$ to $(c)$, the fact that if $k < \ceil{T}$ then $(N - k) > \ceil{T}$ is utilized. Transition from $(c)$ to $(d)$ follows from \eqref{eq:ants1_upper}, while $(e)$ follows from \noteref{note:prior}. In going from $(e)$ to $(f)$, a positive term is dropped. When the target's emitted signal power is large, the term $\text{Pr}\left\{\sum_{i=1}^{\ceil{T}}I_{[i]}  \geq  T|H_1  \right\} \rightarrow 1$. Therefore, when the upper threshold \eqref{eq:upper} is satisfied, ANTS is lower bounded by $(N-\ceil{T})r$.
 
 Next, consider the case when the lower threshold \eqref{eq:lower} is satisfied. Define 
 \begin{align}
k^{\dagger} \!\! = \!\! \min_{1\leq k \leq N} \left\{\sum_{i=1}^{k}I_{[i]}  \leq T - (N-k)    \right\}.
\label{eq:ants1_lower}
\end{align}
The ANTS is given by 
\begin{eqnarray}
\nonumber N^{\ast}\!\!\!\! &=& \mathbb{E}[N - k^{\dagger}] = \sum_{k=1}^{N} (N-k) \text{Pr}(k^{\dagger} = k) \\
\nonumber &\stackrel{(g)}=&\!\!\!\! \sum_{k=1}^{N - \ceil{T} -1} (N-k) \text{Pr}(k^{\dagger} = k) +  \ceil{T}\text{Pr}(k^{\dagger} = N - \ceil{T}) \\ \nonumber && + \sum_{k=N - \ceil{T}+1}^{N} (N-k) \text{Pr}(k^{\dagger} = k) \\ 
\nonumber &\stackrel{(h)}\geq&\!\!\!\! \sum_{k=1}^{N - \ceil{T} -1} (N-k) \text{Pr}(k^{\dagger} = k) +  \ceil{T}\text{Pr}(k^{\dagger} = N - \ceil{T}) \\
\nonumber &\stackrel{(i)}\geq& \ceil{T} \text{Pr}(k^{\dagger} \leq (N - \ceil{T}))  \\
\nonumber &\stackrel{(j)}=& \ceil{T} \text{Pr}\left\{\sum_{i=1}^{N- \ceil{T}} I_{[i]}  \leq T  -  [N - (N- \ceil{T})] \right\} \\
\nonumber &=& \ceil{T} \text{Pr}\left\{\sum_{i=1}^{N- \ceil{T}} I_{[i]}  \leq 0 \right\} \\
\nonumber &\stackrel{(k)}=& \ceil{T} \left[ \text{Pr}\left\{\sum_{i=1}^{N- \ceil{T}} I_{[i]}  \leq 0|H_1 \right\}  r \right. \\ \nonumber && \left. +   \text{Pr}\left\{\sum_{i=1}^{N- \ceil{T}} I_{[i]}  \leq 0|H_0 \right\}(1-r) \right] \\
\nonumber &\stackrel{(l)}\geq& \ceil{T} \text{Pr}\left\{\sum_{i=1}^{N- \ceil{T}} I_{[i]}  \leq 0|H_0 \right\}(1-r).
 \label{eq:ants2}
\end{eqnarray}
In going from $(g)$ to $(h)$, a positive term is dropped. The lower bound in going from $(h)$ to $(i)$ utilizes the fact that if $k \leq N - \ceil{T}$ then $(N - k) \geq \ceil{T}$. Transition from $(i)$ to $(j)$ follows from \eqref{eq:ants1_lower}, while $(k)$ follows from \noteref{note:prior}. In going from $(k)$ to $(l)$, a positive term is dropped. In the absence of a target, it is easy to see that ANTS is lower bounded by $\ceil{T}(1-r)$. Since the two events \eqref{eq:ants1_upper} and \eqref{eq:ants1_lower} are disjoint, when the likelihood of a target present is $0.5$, the ANTS $\geq N/2$. When the target's emitted signal power is large, and when the likelihood of the target present (or absent) is very high, simulation results show that ANTS approaches $(N-1)$.

\section{Numerical results and discussion} \label{sec:simulations}
Under hypothesis $H_1$, $N$ mutually independent sensor locations are generated during each Monte Carlo trial according to the uniform distribution within the ROI according to \eqref{eq:pdf_location}. The $i^{\text{th}}$ local sensor observes $z_i$ and seeks to distinguish between $H_0: z_i = w_i$ and $H_1: z_i = s_i + w_i$. Towards this end, it generates a ``1'' or a ``0'' based on whether the observation $z_i$ exceeds the local decision threshold $\tau$, which can be calculated using \eqref{eq:local_pfa} for a given false alarm rate $p_{\text{fa}}$. The system level threshold $T$ is calculated using \eqref{eq:fusion center_T} for a given $P_{\text{fa}}$ and $p_{\text{fa}}$. A system level detection is declared if either \eqref{eq:upper} or \eqref{eq:lower} is satisfied; otherwise, a missed detection is declared. The ANTS is computed over $10^5$ Monte Carlo trials. For all experiments, the power decay factor $\alpha = 0.02$, while the signal decay exponent $n = 2$. The system-level $P_{\text{fa}}$ and local sensor-level $p_{\text{fa}}$ are both set at $10^{-3}$. 

The ordering-based counting rule achieves the same detection probability as that of the original counting rule detector \cite{Niu2004}. This can be experimentally verified by evaluating the frequency of correct detections during all the Monte Carlo instantiations which gives an estimate of the system level detection probability $P_{\text{d}}$, however, it can be straightforwardly ascertained by inspecting the upper and lower thresholds in \eqref{eq:upper} and \eqref{eq:lower}, respectively. 

\begin{figure}
\centering
\includegraphics[height=3.1in,width=3.5in]{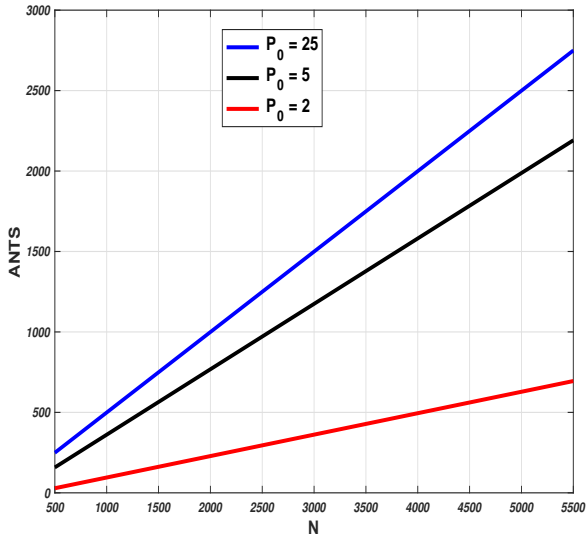}
\caption{ANTS versus $N$ for different values of signal powers.}
\label{fig:ants_n_power}
\end{figure}

The plot of ANTS versus $N$ for different values of signal power $P_0$ is shown in \figref{fig:ants_n_power}.  As can be seen from \eqref{eq:upper} and \eqref{eq:lower}, the ordering-based counting rule achieves the same detection performance as that of the original counting rule detector with far fewer sensor transmissions. The savings approaches $N/2$ for increasing signal powers emitted by the target. Even when the signal power is not very high, savings in transmission is significant. 
\begin{figure}
\centering
\includegraphics[height=3.1in,width=3.5in]{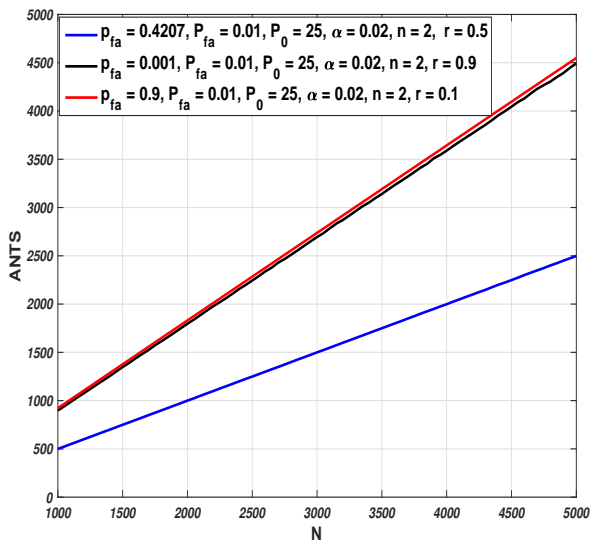}
\caption{ANTS vs. $N$ for different local sensor false alarm rates.}
\label{fig:ants_N_p}
\end{figure}

The plot of ANTS versus $N$ for different values of local sensor false alarm rates and likelihood of target's presence is shown in \figref{fig:ants_N_p}. When the local sensor $p_{\text{fa}}$ is very small, the system-level threshold $T$ takes a very small value. This results in fewer observations to make the correct detection when the likelihood $r$ of the target is high, leading to significant savings in transmission. Similarly, when the local $p_{\text{fa}}$ is large, the system-level threshold $T$ takes a higher value.  This results in fewer observations to make the correct detection when the likelihood of the target is low, again leading to significant savings. In fact, as seen in \figref{fig:ants_N_p} (red and black lines), in these extreme cases ANTS approaches $(N-1)$. When the likelihoods of the target absent or target present are the same, the local sensor $p_{\text{fa}}$ has no impact on the savings as long as the emitted power is very high, leading to an average $N/2$ savings in transmission (blue line in \figref{fig:ants_N_p} which is the same as the blue line in \figref{fig:ants_n_power}).

\section{Concluding remarks} \label{sec:conclusion}
This paper proposes a method for energy-efficient distributed detection of a stationary target in a wireless sensor network comprising a large number $N$ of randomly deployed sensors making observations of the signal emitted by the target. The local decisions of the sensors are transmitted in binary form to a fusion center, where the number of local-sensor detections are counted and compared to a threshold to make a global inference. In previous work, this mechanism was referred to as the counting rule detector, which requires the binary decisions of all the $N$ local sensors to make a global decision. In this paper, the energy efficiency of the counting rule detector is significantly improved without compromising on its detection probability: a local sensor transmits its decision in binary form to the fusion center at a time inversely proportional to a function of the power of its received signal. Then when sufficient evidence is collected at the fusion center, the transmissions are stopped. For large $N$ and when the likelihood of the target's presence (or, absence) is $0.5$, the average number of sensor transmissions saved approaches $N/2$, achieving the same detection probability as that of the counting rule detector. Numerical results show very high savings when the likelihood of the target present (or absent) is very high. The method proposed in the paper exploits the benefits (in terms of energy efficiency) offered by the counting rule and ordering sensor transmissions to significantly improve the energy efficiency of the WSNs.


\bibliographystyle{IEEEtran}
\bibliography{IEEEabrv,order}
\raggedbottom


\end{document}